# Concurrently Non-Malleable Zero Knowledge in the Authenticated Public-Key Model

Yi Deng[†], Giovanni Di Crescenzo[§], Dongdai Lin[‡]

[†] [‡] The state key laboratory of information security,Institute of software,
Chinese Academy of sciences, Beijing, 100080, China

[§] Telcordia Technologies, Piscataway, NJ, USA

Email: {ydeng,ddlin}@is.iscas.ac.cn , giovanni@research.telcordia.com

First version: May 30, 2006
Updated: September 8, 2006

**Abstract**

We consider a type of zero-knowledge protocols that are of interest for their practical applications within networks like the Internet: efficient zero-knowledge arguments of knowledge that remain secure against concurrent man-in-the-middle attacks. As negative results in the area of concurrent non-malleable zero-knowledge imply that protocols in the standard setting (i.e., under no setup assumptions) can only be given for trivial languages, researchers have studied such protocols in models with setup assumptions, such as the common reference string (CRS) model. This model assumes that a reference string is honestly created at the beginning of all interactions and later available to all parties (an assumption that is satisfied, for instance, in the presence of a trusted party).

A growing area of research in Cryptography is that of reducing the setup assumptions under which certain cryptographic protocols can be realized. In an effort to reduce the setup assumptions required for efficient zero-knowledge arguments of knowledge that remain secure against concurrent man-in-the-middle attacks, we consider a model, which we call the Authenticated Public-Key (APK) model. The APK model seems to significantly reduce the setup assumptions made by the CRS model (as no trusted party or honest execution of a centralized algorithm are required), and can be seen as a slightly stronger variation of the Bare Public-Key (BPK) model from [8, 30], and a weaker variation of the registered public-key model used in [3]. We then define and study man-in-the-middle attacks in the APK model. Our main result is a constant-round concurrent non-malleable zero-knowledge argument of knowledge for any polynomial-time relation (associated to a language in $\mathcal{NP}$), under the (minimal) assumption of the existence of a one-way function family. We also show time-efficient instantiations of our protocol, in which the transformation from a 3-round honest-verifier zero-knowledge argument of knowledge to a 4-round concurrently non-malleable zero-knowledge argument of knowledge for the same relation incurs only $\mathcal{O}(1)$ (precisely, a *small* constant) additional modular exponentiations, based on known number-theoretic assumptions. Furthermore, the APK model is motivated by the consideration of some man-in-the-middle attacks in models with setup assumptions that had not been considered previously and might be of independent interest.

We also note a negative result with respect to further reducing the setup assumptions of our protocol to those in the (unauthenticated) BPK model, by showing that concurrently non-malleable zero-knowledge arguments of knowledge in the BPK model are only possible for trivial languages.

## 1 Introduction

Zero-knowledge protocols, first introduced in [24], have received a significant amount of attention from the research community because of their useful applications to several cryptographic protocols in a variety of settings. As such protocols are often deployed in distributed and asynchronous networks like the Internet, the research on these protocols is moving towards extending the security properties of (stand-alone) zero-knowledge protocols to models with multiple parties, asynchronous message delivery, and adversarial modification to exchanged messages.

In particular, the notion of concurrent zero-knowledge, first studied by [18], extends the zero-knowledge security notion to the case where multiple concurrent executions of the same protocol take place and a malicious adversary may control the scheduling of the messages and corrupt multiple provers or verifiers in order to violate the soundness or zero-knowledge properties (respectively). Unfortunately, concurrent zero-knowledge with black-box simulation requires a logarithmic number of rounds for languages outside $\mathcal{BPP}$ [9] and therefore their round-complexity is not efficient. In the Common Reference String model, in [11] it is showed that 3-round and time-efficient concurrent zero knowledge can be achieved. Surprisingly, using non-black-box techniques, Barak [1] constructed a constant round non-black-box bounded concurrent zero knowledge protocol whose time-complexity however is not efficient.

The concept of non-malleable zero knowledge was put forward in [17]. The issue of malleability arises in the so-called man-in-the-middle setting, in which the adversary plays the role of the verifier in several proofs (left interactions) and at the same time acts as the prover in some other proofs (right interactions), having full control over the scheduling of the messages between parties. The serious problem in such scenario is that the information obtained from left interactions may help the adversary to cheat the verifier in one of the right interactions (malleability). A zero-knowledge protocol is considered non-malleable if it is immune against such problem. In [17], the authors give a $\mathcal{O}(\texttt{log}n)$-round non-malleable concurrent zero-knowledge protocol in which the adversary interacts with only one prover. Achieving non-malleability non-interactively in the common random string model was studied in [12] and [37]. In [2], a constant-round coin-tossing protocol assuming the existence of hash functions that are collision-resistant against subexponential-time adversaries was presented, which can be used to transform non-malleable non-interactive zero-knowledge in the shared random string model into interactive non-malleable zero-knowledge in the plain model. A new constant-round non-malleable ZK with minimum assumptions was present in [34], but it failed to be extended to the concurrent model. Note that in [2] and [34] non-black-box techniques were used, and thus the resulting protocols are very inefficient. As showed in [29], it is impossible to achieve concurrent non-malleability without set-up assumption. Several works on this issue show the feasibility to achieve concurrent non-malleability efficiently in the common reference string model. In [20] Garay et al. defined the so-called $\Omega$-protocol, a variant of $\Sigma$-protocol with straight line extractor, and then they show a technique to transform the $\Omega$-protocol to a concurrently non-malleable ZK protocol. Gennaro [21] introduced multi-trapdoor commitments and presented a very efficient ZK protocol enjoying concurrent non-malleability.

A growing area of research in Cryptography is that of reducing the setup assumptions under which certain cryptographic protocols can be realized. In an effort to reduce the setup assumptions required for efficient zero-knowledge arguments of knowledge that remain secure against concurrent man-in-the-middle attacks, we study the concurrent non-malleability in a weak model with very relaxed set-up assumption, the Bare Public-Key (BPK) model of [8]. Comparing with some previous model such as common reference string model and the preprocessing model considered, for instance, in [14], this model seems to significantly reduce the set-up assumptions: It just assumes that each verifier deposits a public key $pk$ in a public file before any interaction with the prover begins, with no need of trusted parties. Since its introduction, many papers focusing on resettable zero knowledge in this model appeared in recent years, but they do not address man-in-the-middle attacks. More specifically, we consider a new variation of this model, which we call the Authenticated Public-Key (APK) model. The APK model is a stronger variation of the BPK model from [8, 30], and can also be seen as a weaker variation of the registered public-key (RPK) model used in [3], and still seems to significantly reduce the setup assumptions made by the RPK or CRS models (as no trusted party or honest execution of a centralized algorithm are required; see detailed discussion in Section 2).

**Our results.** In this paper we study the concurrent non-malleability in the APK model and in BPK model, our results are:

*concurrently non-malleability ZK in APK model based on one-way functions.* This is our main result. In particular, we construct a 5-round concurrently non-malleable zero-knowledge argument of knowledge for any polynomial-time relation (associated to an $\mathcal{NP}$ language), under the (minimal) assumption of the existence of a one-way function family.

*round-optimal and time-efficient instantiations in APK model based on specific number-theoretic assumptions.* We also show two efficient instantiations of our protocol, in which the transformation from a 3-round honest-verifier zero-knowledge argument of knowledge to a 4-round concurrently non-malleable zero-knowledge argument of knowledge incurs only $\mathcal{O}(1)$ (a *small* constant) additional modular exponentiations, based on known number-theoretic assumptions.

*Negative results in BPK model.* We note a negative result with respect to further reducing the setup assumptions of our protocol to the BPK model, by showing that concurrently non-malleable zero-knowledge arguments of knowledge in the (unauthenticated) BPK model are only possible for trivial languages. This shows that the setup assumption behind our main result is essentially optimal.

Our study of the APK model is also partially motivated by the consideration of some man-in-the-middle attacks in models with setup assumptions that had apparently not been considered previously and might be of independent interest. Specifically, we allow a man-in-the-middle to modify not only the communication between two parties (i.e., a prover and a verifier), but also between a party (prover or verifier) and the site of the common setup information, being this a public file with several public keys, a common reference string, etc. (as access to common setup information is itself realized as a communication exchange between the party and the setup information's site).

**Comparison with the recent work [33].** Very recently, Ostrovsky, Persiano, Visconti introduced the notion of concurrent non-malleable witness indistinguishability and implemented a concurrently non-malleable zero knowledge protocol and realized ZK argument of knowledge under general concurrent composition in BPK model. This seems to contradict our impossibility results. However, we argue that the term "BPK" in their paper is actually identical to ours "APK" model: We make the same requirement on BPK model that once the public file is published, the whole file can not be modified by adversary. We think such requirement must involve in authenticated channel in the man-in-the-middle setting if we allow adversaries to control all communication that takes place in the proof stage (including the prover's access to the public file) as considered in this paper. This is the only reason we call the new model "APK" model.

Ignoring the above artificial difference, we note that, compared to OPV's construction for concurrently non-malleable ZK argument, Our construction is far more efficient because we use only black-box technique, and is based on more general intractability assumption (existence of one-way functions). Furthermore, based on some specific number-theoretic assumptions, We can also instantiate our protocol very efficiently: the transformation from a 3-round honest-verifier zero-knowledge argument of knowledge to a 4-round (optimal) concurrently non-malleable zero-knowledge argument of knowledge for the same relation incurs only a *small constant* additional modular exponentiations.

We also note that Ostrovsky et al. address more general issues, e.g., ZK under general concurrent composition, which we do not study here.

## 2 Definitions

In this section we define the authenticated public-key (APK) model for zero-knowledge protocols, and define concurrently non-malleable zero-knowledge arguments in this model. We also recall the background tools of commitment schemes, signature schemes and $\Sigma$-protocols, that will be used in our main construction.

**The APK Model.** In [8], the authors introduced the bare public-key (BPK) model for zero-knowledge protocols. Informally, in this model, protocols are defined in two stages: a preprocessing stage, in which all verifiers post a public key on a public file; and a proof stage, in which provers and verifiers interact and have access to the previously created public file. The term 'bare' in the name of the model refers to the fact that, contrarily to other models, such as the public-key infrastructure model, no certification of the public keys is required. In this model, the authors (and several other papers) presented constant-round resettable zero-knowledge arguments for any language in $\mathcal{NP}$, under appropriate complexity assumptions. In this paper we would like to construct zero-knowledge arguments of knowledge that remain so even under man-in-the-middle attacks. To this purpose, we consider a slightly stronger version of the BPK model, which we call the authenticated public-key (APK) model. Briefly speaking, the APK model only augments the BPK model in that during the proof stage, the provers are guaranteed to have access to the same public file that was determined at the end of the preprocessing stage. This happens regardless of any adversarial activity such as man-in-the-middle attacks (in other words, adversaries in the middle between any prover and the public file's site are prevented to alter the prover's reading of this file). Formally, the authenticated public-key model (APK model) makes the following assumptions.

1. There are two types of entities: provers and verifiers, and the entire interaction between them can be divided into two stages; the first stage is called *preprocessing stage* and only needs to be run by verifiers; at the end of the preprocessing stage, the *proof stage* starts, where any pair of prover and verifier can interact. All algorithms have access to a public file. Provers, verifiers and the public file are defined below.
2. The *public file*, structured as a collection of records, is empty at the beginning and can be modified by the verifiers during the preprocessing stage; the version of the public file $F$ obtained at the end of the preprocessing stage will be used during the proof stage.
3. An (honest) *prover* $P$ is an interactive deterministic polynomial-time algorithm that operates in the proof stage, on input a security parameter $1^n$, an $n$-bit string $x \in L$, an auxiliary input $w$, a public file $F_p$ and a random tape $r_p$, where $L$ is a language in $\mathcal{NP}$.
4. An (honest) *verifier* $V$ is a pair of deterministic polynomial-time algorithms $(V_1, V_2)$, where $V_1$ operates in the preprocessing stage and $V_2$ operates in the proof stage. On input a security parameter $1^n$ and a random tape $r_{v1}$, $V_1$ generates a key pair $(pk, sk)$ and stores $pk$ in the public file. On input $pk, sk$, an $n$-bit string $x$ and a random tape $r_{v2}$, the interactive algorithm $V_2$ performs the interactive protocol with a prover, and outputs "accept $x$" or "reject $x$" at the end of this protocol.
5. There is an authenticated channel between a prover and the public file site. Thus for any prover it holds that $F_p = F$.

*Remarks.* The only difference between the above formal definition and the formal definition of the BPK model from [8, 30] is in item 5, which implies that the public file used by all provers cannot be different from the public file obtained at the end of the preprocessing phase. We note however that all previous papers in the BPK model (starting with [8, 30]) did not study man-in-the-middle attacks, but mostly focused on questions related to concurrent or resettable variants of soundness or zero-knowledge properties. When considering man-in-the-middle attacks, as we do in this paper, we need to consider adversaries that can control all communication that takes place in the proof stage (including the prover's access to the public file) and invoke many different provers in the left interactions (these provers are not aware of the existence of each other). Although not explicitly stated, a similar constraint on the public file has to implicitly hold for each verifier; specifically, first recall that verifiers are defined in two stages; then note that each verifier's proof stage activation uses the same public file (or, more precisely, the same pair $(pk, sk)$) generated by the same verifier's preprocessing stage activation. In other words, each verifier is already assumed to 'remember' (at least part of) the public file between the two stages in the BPK model, and the APK model could be realized by further assuming that also each prover is assumed to remember the (entire) public file between the two stages (thus obviously realizing authenticated access as access to private memory).

*Comparison with the RPK and CRS models.* We note that the APK model can be seen as a weaker version of the RPK model, for registered public-key model from [3], a stronger version of the BPK model, where all parties (i.e., both provers and verifiers) are required to post a public key and a trusted party needs to verify each party's knowledge of the associated secret key in the preprocessing stage. (To see that the RPK model is stronger than the APK model, note that each party in the RPK model can keep state information, such as the content of the entire public file between the preprocessing stage and the proof stage, and therefore authenticated access to the public file is obviously guaranteed by a party accessing its own state information.) In the APK model, however, it is not required that the provers publish a public key and no trusted party is necessary in the preprocessing/registering stage. The model in [3] was used to study secure function evaluation protocols with universal composability properties, where the authors' goal was that of presenting such protocols under setup assumptions weaker than those in the common reference string model. Analogously, in this paper we present concurrently non-malleable arguments in the APK model, under setup assumptions weaker than those in the CRS model (although the goals addressed and techniques used are very different). We remind readers that if the man-in-the-middle adversary is allowed to control the honest party's access to the common random string, the CRS model also should assume an authenticated channel between all parties and the site where the common reference string is available (so that all parties have a secure access to the common reference string).

**Concurrently Non-Malleable Arguments of Knowledge in the APK Model.** We start with some basic definitions and then define the requirements of completeness, concurrent zero-knowledge, extraction, and view simulatability.

We say that function $f(n)$ is *negligible* if for every polynomial $q(n)$ there exists a positive integer $n_0$ such that for all $n \geq n_0$, it holds that $f(n) \leq 1/q(n)$. If $L$ is a language in $\mathcal{NP}$, we define the *associated relation* as the relation $R_L = \{(x,w) \,|\, x \in L; w \text{ is a witness for '} x \in L\text{'}\}$. Conversely, let $R$ be a relation. We define the *domain* of $R$ as the language dom $R = \{x \,|\, \exists w \text{ such that } (x,w) \in R\}$. If $R$ is a polynomial-time relation (i.e., a relation for which there exists a polynomial-time algorithm deciding whether an input pair belongs to it), we define the *associated language* $L_R$ as the domain of $R$. The definition of class $\mathcal{NP}$ implies that for every language $L$ in $\mathcal{NP}$, its associated relation $R_L$ is polynomial-time.

*Completeness and Concurrent Zero-knowledge.* The definition of completeness and concurrent zero-knowledge of argument systems in the APK model are immediate adaptations of the analogous definitions in the BPK model, which, in turn, are adaptations of the analogous definitions in the standard model.

**Definition 2.1** Let $L$ be a language $L$ in $\mathcal{NP}$ and let $R_L$ be its associated relation; also, let $P$ and $V$ be a prover and a verifier, respectively, in the APK model.

We say that pair $(P,V)$ satisfies the *completeness* requirement if, after public file $F$ has been generated during the preprocessing phase, where $F$ contains pair $(pk, sk)$ generated by $V_1$, for all $n$-bit strings $x \in L$ and any $w$ such that $(x,w) \in R_L$, the probability that in the proof stage $V$ interacting with $P$ on input $y$, outputs "reject" is negligible in $n$.

We say that pair $(P,V)$ satisfies the *concurrent (black-box) zero-knowledge* requirement if there exists a probabilistic polynomial-time algorithm $S$ such that for any probabilistic polynomial-time algorithm $V^*$, for any polynomials $s, t$, for any $x_i \in L$, the length of $x_i$ is $n$, $i = 1, ..., s(n)$, $V^*$ runs in at most $t(n)$ steps and the following two distributions are indistinguishable:
  1. the output of $V^*$ that firstly generates $F$ with $s(n)$ entries and interacts concurrently with $s^2$ instances of the honest prover: $P(x_i, w_i, pk_j, F)$, $1 \leq i, j \leq s(n)$, where each instance uses an independent random string, $w_i$ is a witness for $x_i \in L$, and $pk_j$ is the $j$-th entry registered by $V$ in $F$.
  2. the output of $S$ with on input $x_1, ... x_{s(n)}$.

*The adversary and its man-in-the-middle attacks in the APK model.* Let $s$ be a positive polynomial. The adversary we consider, called a *man-in-the-middle in the APK model*, and denoted as $\mathcal{A}$, is a polynomial-time algorithm that can act both as a prover and as a verifier, both in the preprocessing stage and in the proof stage. Specifically, in the preprocessing stage, $\mathcal{A}$ can act as a verifier and register up to a polynomial number of public keys. In the proof stage, $\mathcal{A}$ can concurrently interact with provers in *left interactions*, where it plays the role of a verifier and uses a public key posted by $\mathcal{A}$ itself or by a (real) verifier; or $\mathcal{A}$ can interact with verifiers in *right interactions*, where it plays the role of a prover. The total number of left and right sessions is at most $s(n)$, where $n$ is a security parameter (also set to be equal to the length of the instances of the proved statements).

More formally, $\mathcal{A}$'s attack is executed as follows:

- *Preprocessing stage:* $\mathcal{A}$ takes as inputs $1^n$ and a random string $r$ and registers a polynomial number of public keys (we denote this polynomial as $t$). All honest verifiers register their public keys (for the sake of simplicity of notation, we also assume the number of the honest verifier's public keys are at most $t(n)$). The public keys registered by $\mathcal{A}$ and all honest verifiers form a public file $F$.

- *Proof stage:* $\mathcal{A}$ continues its execution by taking $F$ as additional input, and may start a polynomial number of left interactions or right interactions (for the sake of simplicity of notation, we assume that this number is at most $s(n)$).

  At any time during this stage $\mathcal{A}$ can do one of the following four actions:

  1) deliver to $V$ a message for an already started right session; 2) deliver to $P$ a message for an already started left session; 3) start a new left session, by choosing a new statement '$y_j \in L$', a public key $pk_j$ from $F$ (which was previously generated by either a honest verifier or even by $\mathcal{A}$ itself), and a prover $P$, and send a special message 'Start left session with inputs $y_j, pk_j$' to $P$, who is given as additional input $w_i$ such that $R_L(y_j, w_j) = 1$ and has access to the (not modified) public file $F$; 4) start a new right session, by choosing a new statement '$x_i \in L$', a verifier $V$ and a public key $pk_i$ from $F$ which was previously generated by $V$, and send a special message 'Start right session with inputs $x_i, pk_i$' to $V$, who is given as additional input the secret key $sk_i$ associated with $pk_i$, and a new random string as random tape; 5) output a special 'end attack' symbol, where we assume that this symbol is returned within time polynomial in $n$.

  In all cases, the prover $P$'s (or verifier $V$'s) reply (if any) to $\mathcal{A}$'s message is immediately delivered to $\mathcal{A}$.

The adversary's goal is that of completing a right session for which the verifier accepts even if the adversary does not know a witness for the associated statement.

Let $tr_{right}(y)$ (resp., $tr_{left}(x)$) be the transcript of an session generated by $\mathcal{A}$ and $V$ (resp., $\mathcal{A}$ and $P$) on common input $x$ (resp., $y$) under one public key adaptively chosen by $\mathcal{A}$, that was previously registered by a verifier $V$ on the public file $F$. Also, we denote by $\lambda \leftarrow \Lambda(x, y, \ldots)$ the process of running the probabilistic algorithm $\Lambda$ on input $(x, y, \ldots)$, and obtaining $\lambda$ as output.

We consider the probability that $V$ accepts the statement '$x_i \in L$' in a right session and the transcript of this interaction is different from any transcript in the left interactions. This probability is denoted as $p^i_\mathcal{A}(x_i)$ and can be formally defined as equal to:

$$Prob[(pk, sk) \leftarrow V(1^n); \mathcal{A}^{P_1(y_1),\ldots P_s(y_s), V_1(x_1),\ldots V_s(x_s)}(pk, x_i), V(pk, sk, x_i)] = 1 \wedge tr_{right}(x_i) \notin Q]$$

Where $V$ is one of the verifiers $V'_i$s, the statements $y_1, ..., y_s, x_1, ..., x_i, ..., x_s$ are adaptively chosen by $\mathcal{A}$, and $Q = \{tr_{left}(y_j) : 1 \leq j \leq s\}$.

To define the extraction requirement, we note that the extractor acts in two phases: first, it interacts with the adversary $\mathcal{A}$, by returning an 'extended' transcript $extr_{right}(x_i)$ related to a particular right session. Specifically, $extr_{right}(x_i)$ contains all messages exchanged between $\mathcal{A}$ and the extractor within a particular right interaction where the statement '$x_i \in L$' is being proved. (Note that this is an extended version of $tr_{right}(x_i)$ as it may include messages exchanged across different executions of the same session due to multiple rewindings of $\mathcal{A}$ done by the extractor.) Second, a predicate $\rho$ tries to extract a witness from $extr_{right}(x_i)$, and is successful whenever $p^i_\mathcal{A}(x_i)$ is large enough. We stress that restricting the extractor in the APK model to return a witness based only on the particular transcript $extr_{right}(x_i)$ (rather than, say, based on transcripts of all $s(n)$ sessions) is necessary due to the separation results in the BPK model of [30]; in this latter paper, messages from other sessions may be used as witnesses for the current session.

We are finally ready to define the two requirements of extraction and simulatability of $\mathcal{A}$'s view, and we can then define concurrently non-malleable zero knowledge argument of knowledge in the APK model.

**Definition 2.2** Let $L$ be a language $L$ in $\mathcal{NP}$ and let $R_L$ be its associated relation; let $P$ and $V$ be a prover and a verifier, respectively, in the APK model, and let $\mu : \mathbb{N} \to [0,1]$ be a function (knowledge error).

We say that pair $(P,V)$ is a *(black-box) concurrently non-malleable zero knowledge argument of knowledge (with knowledge error $\mu$) for relation $R_L$* if it satisfies the following requirements:

1. *Completeness and Concurrent Zero Knowledge*, as in Definition 2.1.

2. *Extraction:* For every polynomial $s$ and every probabilistic polynomial-time man-in-the-middle adversary $\mathcal{A}$ that engages in at most $s$ sessions in each part of interactions (left or right interactions), there exists a probabilistic polynomial time knowledge extractor $\mathcal{E}$ and a polynomial-time predicate $\rho$, such that for all $i$, $1 \leq i \leq s$, if $p^i_{\mathcal{A}}(x_i) > \mu(n)$ then $\mathcal{E}$, with access to $\mathcal{A}$, returns $extr_{right}(x_i)$ and $\rho(x_i, pk, extr_{right})$ returns $w_i$ such that $(x_i, w_i) \in R_L$ with probability differing from $p^i_{\mathcal{A}}(x_i) - \mu(n)$ only by a negligible amount.

3. *Simulatability of $\mathcal{A}$'s View:* There exists a probabilistic polynomial-time simulator $\mathcal{M} = [\mathcal{M}_P, \mathcal{M}_v]$ such that $\mathcal{A}$'s view $\text{View}[\mathcal{A}, P, V]$ in the real man-in-the-middle setting is computational indistinguishable from the view $\text{View}[\mathcal{A}, \mathcal{M}_P, \mathcal{M}_V]$ simulated by $\mathcal{M}$. Here, $\text{View}[\mathcal{A}, X, Y]$ is formally defined as the sequence of $\mathcal{A}$'s random coins, all common inputs and public keys, and the transcripts of all left and right interactions between $\mathcal{A}$ and any provers or verifiers executed using algorithms $X$ and $Y$, respectively.

**Background tools.** We recall definitions of known tools, as commitment schemes, signatures, and $\Sigma$-protocols, which will be useful in our main construction.

*Commitment schemes.* A commitment scheme is a two-phase two-party (the sender $S$ and the receiver $R$) interactive protocol with the following properties: 1) correctness: at the end of the second phase, $R$ obtains the value committed by $S$ during the first phase; 2) hiding: commitment keys associated to two committed values are computationally indistinguishable by every probabilistic polynomial-time (possibly malicious) $R^*$; 3) binding: after having committed to a value $m$ during the first phase, any probabilistic polynomial-time (possibly malicious) sender $S^*$ cannot open this commitment to another value $m' \neq m$ except with negligible probability. We will mainly use commitment schemes where the first phase consists of 2 messages: one preliminary message sent by $R$, and one message sent by $S$; and the second phase consists of 1 message sent by $S$. Such schemes can be constructed assuming the existence of any one-way function families (using the scheme from [31] and the result from [27]) or under number-theoretic assumptions (e.g., the scheme from [35]). Assuming the existence of one-way permutation families, a well-known construction of a commitment scheme (see, e.g. [22]) can be given where the first phase only consists of a single message from $S$.

*Signature schemes.* A signature scheme is a triplet, $(KG, Sig, Ver)$, of probabilistic polynomial-time algorithms. $KG$ is the key-generation algorithm that on input the security parameter $1^n$ generates a key pair $(sig\_k, ver\_k)$. On input the signing key $sig\_k$ and a massage $m$ algorithm $Sig$ outputs a signature $\sigma$ for $m$, i.e, $\sigma = Sig(sig\_k, m)$. Given $(ver\_k)$, a message $m$ and a string $\sigma'$, algorithm $Ver$ outputs 1 if $\sigma'$ is a valid signature for $m$, i.e, $Ver(ver\_k, \sigma', m) = 1$, otherwise outputs 0.

We will use two properties of many signature schemes in the literature.

We say a signature scheme is a *one-time strong* signature scheme, if for all probabilistic polynomial-time adversary, after received one valid signature of a message of its choice by querying the signing oracle, this adversary can not produce a different valid signature on any message even including the message he queried. We can get such signature schemes based on assumption of existence of one-way function families, for instance, from [19].

We consider signature schemes that satisfy *existential unforgeability against adaptive chosen message attack* (first defined in [25]). The security of such schemes is formalized in a scenario in which an adversary requests the signing oracle for signatures on a polynomial (in the security parameter $n$) number $s(n)$ of messages at its choice. If for all polynomial $s$ there is no probabilistic polynomial-time adversary which forges a valid signature on a message different from all messages he queried, we say that the signature scheme is secure against adaptive chosen message attack. As showed in [36], One-way functions are sufficient for this type of signature.

$\Sigma$-*protocols.* These protocols are defined as 3-round public-coin proofs of knowledge with some nice properties: 1) *special soundness.* Let $(a, e, z)$ be the three messages exchanged by prover $P$ and verifier $V$ in a session. From any statement $x$ and any pair of accepting transcripts $(a, e, z)$ and $(a, e', z')$ where $e \neq e'$, one can efficiently compute $w$ such that $(x, w) \in R$. 2) *Special honest-verifier zero-knowledge:* given the second message $e$ and the statement $x$, we can compute an accepting transcript of form $(a, e, z)$ that is computational indistinguishable from the real transcript between $P$ and the honest $V$.

Many known efficient protocols, such as those in [26] and [38], are $\Sigma$-protocols. Furthermore, there is a $\Sigma$-protocol for the language of Hamiltonian Graphs [7], assuming that one-way permutation families exists; if the commitment scheme used by the protocol in [7] is implemented using the scheme in [31] from any pseudo-random generator family, then the assumption can be reduced to the existence of one-way function families, at the cost of adding one preliminary message from the verifier. (See previous discussion about construction of commitment schemes.) We will refer to this modified protocol as a *4-round $\Sigma$-protocol*, and we will use the fact that any language in $\mathcal{NP}$ admits a 4-round $\Sigma$-protocol under the existence of any one-way function family (or under an appropriate number-theoretic assumption), or a $\Sigma$-protocol under the existence of any one-way permutation family. We will also use *partially-witness-independent* $\Sigma$-protocols, where only the last message in these protocols depends on the witness of the proved statement, while all other messages only depend on (an upper bound on) the length of any such witness. Many $\Sigma$-protocols (including [7] for all of $\mathcal{NP}$) are partially-witness-independent.

Interestingly, $\Sigma$-protocols can be composed to proving the OR of atomic statements, as shown in [13, 10]. Specifically, given two protocols $\Sigma_0, \Sigma_1$ for two relationships $R_0$, $R_1$, respectively, we can construct a $\Sigma_{OR}$-protocol for the following relationship efficiently: $R_{OR} = ((x_0, x_1), y) : (x_0, y) \in R_0 or (x_1, y) \in R_1$, as follows. Let $(x_b, y) \in R_b$ and $y$ is the private input of $P$. $P$ computes $a_b$ according the protocol $\Sigma_b$ using $(x_b, y)$. $P$ chooses $e_{1-b}$ and feeds the simulator $M$ guaranteed by $\Sigma_{1-b}$ with $e_{1-b}, x_{1-b}$, runs it and gets the output $(a_{1-b}, e_{1-b}, z_{1-b})$. $P$ sends $a_b, a_{1-b}$ to $V$ in first step. In second step, $V$ picks $e$ randomly and sends it to $P$. Last, $P$ sets $e_b = e \oplus e_{1-b}$, and computes the last message $z_b$ to the challenge $e_b$ using $x_b, y$ as witness according the protocol $\Sigma_b$. $P$ sends $e_b$, $e_{1-b}, z_b$) and $e_{1-b}, z_{1-b}$ to $V$. $V$ checks $e = e_b \oplus e_{1-b}$, and the two transcripts $(a_b, e_b, z_b)$ and $(a_{1-b}, e_{1-b}, z_{1-b})$ are accepting. The resulting protocol turns out to be witness indistinguishable: the verifier can not tell which witness the prover used from a transcript of a session.

In our construction the verifier executes a $\Sigma_{OR}$-protocol to prove the knowledge of one of the secret keys corresponding to his public key. Furthermore, as required in [16], we need a *partial-witness-independence* property from this protocol: the message sent at its first round should have distribution independent from any witness for the statement to be proved. We can obtain such a protocol using [38], [13].

# 3 Constant-Round Concurrently Non-Malleable Zero-Knowledge Arguments of Knowledge in the APK Model

In this section we present our main result: under general complexity assumptions, we present a constant-round concurrently non-malleable zero-knowledge argument of knowledge for any polynomial-time relation in the APK model. We describe our result as a transformation that, under general complexity assumptions, applies to any $\Sigma$-protocol (that is, an argument of knowledge with negligible knowledge error that satisfies special soundness and special honest-verifier zero-knowledge). Formally, we obtain the following

**Theorem 3.1** Let $L$ be a language in $\mathcal{NP}$, let $R_L$ be the associated polynomial-time relation, and assume that there exists a $\Sigma$-protocol for $R_L$. In the APK model, if there exist one-way function families, then there exists a constant-round (black-box) concurrently non-malleable zero-knowledge argument of knowledge for $R_L$.

An important consequence of this theorem is that under the existence of one-way function families, we obtain a constant-round (black-box) concurrently non-malleable zero-knowledge argument of knowledge for $R_L$, for any language $L$ in $\mathcal{NP}$.

*Remarks on complexity assumptions, round and time efficiency.* We note that the complexity assumption sufficient for our transformation is the weakest possible as it is also necessary for non-trivial relations, due to one of the many consequences of the main result in [28]. The above statement for our main result focuses on strongest generality with respect to complexity assumptions and neglects round-optimality and time-efficiency of prover and verifier. However, we note that our transformation takes 5 rounds when implemented using arbitrary one-way function families, and can be implemented in only 4 rounds (which is optimal, due to a result in [23]), when implemented using one-way permutation families (see also discussion at the end of Section 2 for the assumptions and rounds required to implement $\Sigma$-protocols for $\mathcal{NP}$-complete languages). Furthermore, we note that time-efficient instantiations of our transformation, which incur only $\mathcal{O}(1)$ (a *small* constant) additional modular exponentiations, are possible under appropriate number-theoretic intractability assumptions, as we show in Section 3.1.

*Informal description of the difficulties solved by our transformation.* The natural starting point for our transformation is the concurrently non-malleable zero-knowledge argument of knowledge in the CRS model from [20]. Informally speaking, this protocol goes as follows: the prover first generates a key pair $(sk', vk')$ of a one-time strong signature scheme, then sends $vk'$ and proves (using a $\Sigma_{OR}$-protocol $\Pi_p$) that either he knows a witness for the statement to be proved or he knows a valid signature of $vk'$ under a signature verification key $vk$ from the common reference string. In the last step, the prover signs on the whole transcript of this session using $sk'$ and sends the signature to the verifier. Furthermore, the simulator for this protocol uses its knowledge of the secret key associated with the signature verification key $vk$ in the common reference string.

A first way to adjust this protocol so that it might work in the APK model is as follows. Instead of taking $vk$ from the reference string (which is not available in the APK model), we require the verifier to choose $vk$, and to give to the prover a witness-indistinguishable proof that the verifier knows the secret key $sk$ associated with $vk$ (note that omitting this latter proof might make it easier for cheating verifiers to violate the zero-knowledge requirement). Unfortunately, several standard attempts to present a proof for this protocol actually fail, one major problem being in the fact that an algorithm trying to use a cheating prover to break the signature scheme seems to need itself knowledge of the signature secret key in order to be able to use the cheating prover's power. (This does not lead to a contradiction of the security of the signature scheme.)

A second adjustment to this protocol is as follows: the verifier chooses two signature verification keys $vk_0, vk_1$ (rather than one), and proves knowledge of at least one of the two associated secret keys to the prover. Analogously, the prover proves knowledge either of a witness for the statement to be proved or of a valid signature of $vk'$ under any one of the two signature verification keys $vk_0, vk_1$. Unfortunately, even after this additional fix we cannot rule out malleability interactions between the verifier's and the prover's subproofs in this protocol. (A similar situation was detailed in [16], where a specific message schedule was given, and it was showed that a malicious prover could use this schedule and malleability attacks to elude extraction attempts. Unfortunately, a fix based on the solution to this problem proposed by [16] would result in an inefficient protocol that would require $O(1)$ exponentiations for every bit of the security parameter.

Our final fix is that of asking the prover to commit to a random string and prove the knowledge of either a witness for the statement to be proved or a valid signature of $vk'$ under one of the two signature verification keys $vk_0, vk_1$, *where this signature is equal to the value that the prover committed to*. We will show that with this combination of signatures and commitments we avoid the malleability attacks from [16] *efficiently* (i.e., we show instantiations of the overall transformation under appropriate number-theoretic assumptions that only require

---

**The protocol** $(P, V)$

**Security parameter:** $1^n$.
**Common input:** the public file $F$, $n$-bit string $x \in L$, an index $i$ that specifies the $i$-th entry $pk_i = (ver\_k_0, ver\_k_1)$ in $F$, where $(ver\_k_0, ver\_k_1)$ are two verification keys of two signature schemes $(KG_0, Sig_0, Ver_0)$ and $(KG_1, Sig_1, Ver_1)$ that are both secure against adaptive chosen message attack.
**The Prover's private input:** a witness $w$ for $x \in L$.
$V$**'s Private input:** a secret key $sk$ ($sk$ is one of the signing keys corresponding to $(ver\_k_0, ver\_k_1)$, i.e, $sk = sig\_k_0$ or $sig\_k_1$.

$P$ **Step 0:**
compute and send to $V$ the first message of a 4-round and partially-witness-independent $\Sigma_{OR}$-protocol $\Pi_v$ in which $V$ will prove knowledge of $sk$ that is one of the signing keys $(sig\_k_0, sig\_k_1)$;

$V$ **Step 1:**
compute and send to $P$ the second message of the 4-round $\Sigma_{OR}$-protocol $\Pi_v$; compute and send to $P$ the first message $f$ of a 4-round $\Sigma_{OR}$-protocol $\Pi_p$ in which $P$ will prove to $V$ that it knows a witness for statement $x \in L$ or the decommitment $(r, s)$ for $C$, where $s$ is a valid signature of $vk'$ corresponding to $Ver_0$ or $Ver_1$;

$P$ **Step 1:**
1. generate a key pair $(sk', vk')$ for a one-time strong signature scheme $(KG', Sig', Ver')$;
2. choose a random string $s$ of the length of a signature; pick a random string $r$ and compute the commitment key $C = COM(s, r)$;
3. compute and send to $V$ the second message $a$ of the 4-round $\Sigma_{OR}$-protocol $\Pi_p$;
4. send $vk'$, $C$, $a$ and a random string (i.e., the challenge of the $\Sigma$-protocol) as the third message of $\Pi_v$ to $P$.

$V$ **Step 2:**
1. compute the fourth message of protocol $\Pi_v$ according to the challenge sent by $P$ in $P$ **Step 1**; send this message to $P$;
2. send a random challenge $e$ of protocol $\Pi_p$ to $P$;

$P$ **Step 2:**
check whether the transcript of protocol $\Pi_v$ is accepting; if so, compute the last message $z$ of protocol $\Pi_p$; let $tran$ denote the transcript of above interaction (i.e, the whole sequence of messages sent between parties, including $z$); compute the signature $\delta = Sig'(sk', tran)$ and send $z, \delta$ to $V$;

$V$ **Step 3:** accept if and only if $(f, a, e, z)$ is an accepting transcript of $\Pi_p$ and $Ver'(vk', \delta, tran) = 1$.

---

Figure 1: The concurrently non-malleable ZK argument of knowledge in the APK model.

$O(1)$ additional exponentiations) and only using *general complexity assumptions*, such as the existence of one-way function families for the 5-round variant and of one-way permutation families for the 4-round variant. On the other hand, as we allow the man-in-the-middle adversary to register its own public keys, the analysis of security is more involved: our proof of the extraction property of this protocol makes a novel combined use of concurrent scheduling analysis and signature-based simulation arguments.

*proof of theorem 3.1.* The proofs for the properties of *completeness* and *concurrent zero-knowledge* of our protocol are very similar to proofs given in other papers (see, e.g. [20, 16, 39]), and we omit them here for the sake of space. Instead, in the rest of this section we focus on the most interesting property of extraction. (The property *simulatability of $\mathcal{A}$'s view* follows directly from the proof of extraction).

**Extraction.** We now prove that the protocol $(P, V)$ from Figure 1 is an argument of knowledge with negligible knowledge error. Let $\mathcal{A}$ be a probabilistic polynomial-time man-in-the-middle adversary $\mathcal{A}$ that engages in at most $s(n)$ sessions in each type of interactions (left or right interactions), for some polynomial $s$. We would like to show a probabilistic polynomial-time knowledge extractor $\mathcal{E}$ and a polynomial-time predicate $\rho$, such that for all $i$, $1 \leq i \leq s$, if $p^i_{\mathcal{A}}(x_i)$ is not negligible, then $\mathcal{E}$, with access to $\mathcal{A}$, returns $extr_{right}(x_i)$ and $\rho(x_i, pk, extr_{right})$ returns $w_i$ such that $(x_i, w_i) \in R_L$ with probability differing from $p^i_{\mathcal{A}}(x_i)$ only by a negligible amount.

Without loss of generality, we can assume that $\mathcal{A}$ interacts with only one honest verifier $V$ (that is, all right interactions use this specific $V$'s public key), and interacts with many provers under any public keys chosen by $\mathcal{A}$ from the public file $F$ (the extension to the general case with right interactions with multiple honest verifiers is straightforward). Then we can assume that $\mathcal{A}$ engages in at most $s$ left interactions and makes $V$ accept at the end of the $i$-th right session on input statement $x_i \in L$ with a probability $p^i_{\mathcal{A}}(x_i)$ that is not negligible, and the transcript of this session is different from any transcript in the left interactions. For such $\mathcal{A}$, we construct an extractor $\mathcal{E}$ and a polynomial-time predicate that satisfy the extraction requirement in Definition 2.2. The extractor $\mathcal{E}$ is active in both stages: the preprocessing stage and the proof stage.

*The extractor in the preprocessing stage.* On input the security parameter $1^n$, $\mathcal{E}$ generates two key pairs $(sig\_k_0, ver\_k_0)$ and $(sig\_k_1, ver\_k_1)$ for the signature scheme secure against adaptive chosen message attack; then, it registers the public key $pk = (ver\_k_0, ver\_k_1)$ on the public file and keeps $sig\_k_b$ as its secret key (it also stores $sig\_k_{1-b}$, which will be important for the extractor to successfully extract a witness), where $b$ is a random bit selected by $\mathcal{E}$. Then $\mathcal{E}$ runs $\mathcal{A}$'s key generation algorithm and gets its outputs, i.e., a polynomial number of public keys that $\mathcal{A}$ registers on the public file during the preprocessing stage. At the end of this stage, all parties are assumed to obtain the (same) public file.

*The extractor in the proof stage.* We first explain informally a high-level view of algorithm $\mathcal{E}$ in this phase. First of all, upon receiving in a left interaction an accepting conversation of a subprotocol $\Pi_v$ where $\mathcal{A}$ plays as the prover within $\Pi_v$, $\mathcal{E}$ extracts the secret key associated to this execution of $\Pi_v$. Every single one of these extractions is successful due to the extraction properties of the $\Sigma$-protocol $\Pi_v$. We note here that $\mathcal{A}$'s success with high probability to complete executions of $\Pi_v$ does not mean that $\mathcal{A}$ actually knows the secret keys corresponding to those public keys it chooses, as he may learn from right sessions to prove that he knows these secret keys; still, if the probability of $\mathcal{A}$' success is high, $\mathcal{E}$ can extract these secret keys (this is why we can use 'potentially malleable' proofs of knowledge as subprotocol $\Pi_v$). Moreover, all these extractions can be successfully performed by multiple rewindings as in the extractor used in many other papers (e.g., the main protocol in [14]), and since there are at most a polynomial number of such secret keys, the expected number of rewindings is polynomial and this entire process takes at most expected polynomial time. These secret keys allow $\mathcal{E}$ to successfully simulate both the prover in the left interactions (as using a secret key it is possible to compute in polynomial time a witness for the statement proved using subprotocol $\Pi_p$) and the verifier in the right interactions (as $sig\_k_b$ is itself a witness for the statement proved using subprotocol $\Pi_v$) with $\mathcal{A}$ during the proof stage. A correct simulation of the interaction with $\mathcal{A}$ is necessary for $\mathcal{E}$ to later perform one more extraction in correspondence of the $i$-th right session, and thus obtain the desired witness. The hard part is then to prove that $\mathcal{E}$ actually obtains a witness for $R_L$, rather than a signature that $\mathcal{A}$ somehow was able to obtain by some malleability attack; this proofs uses the properties of all three tools of commitments, signatures and partially-witness-independent $\Sigma$-protocols, and is detailed later.

We now proceed more formally with the description of $\mathcal{E}$ in the proof stage, by describing $\mathcal{E}$'s instructions with respect to each message type sent by $\mathcal{A}$:

1. upon receiving *start left session with inputs $y$ (the statement to be proven), $pk$* from $\mathcal{A}$: $\mathcal{E}$ starts a left interaction with inputs $y, pk'$, playing as the prover in '$P$ step 0' (thus sending the first message of the 4-round $\Sigma$-protocol $\Pi_v$).
2. upon receiving *start right session on input $x, pk$* from $\mathcal{A}$: along with this request, $\mathcal{E}$ also receives from $\mathcal{A}$ the message computed according to '$P$ step 0'; then $\mathcal{E}$ computes a message as in '$V$ step 1' and sends it to $\mathcal{A}$;
3. upon receiving a message generated as in '$V$ step 1' (left interaction) on input $y, pk'$: if the secret key corresponding to $pk'$ has not been extracted yet and $pk' \neq pk$, then $\mathcal{E}$ sends to $\mathcal{A}$ a message computed as in '$P$ step 1'; otherwise $pk' = pk$ or $\mathcal{E}$ has already extracted the secret key associated with $pk'$: in the former case, $\mathcal{E}$ flips a random bit $j$, generates a key pair $(sk', vk')$ for the one-time strong signature scheme and computes the signature $\text{sig} = Sig_j(sig\_k_j, vk')$ (note that $\mathcal{E}$ stores both $sig\_k_b$ and $sig\_k_{1-b}$) and computes $c = COM(\text{sig}, r)$ by choosing a random string $r$, and finally executes the protocol $\Pi_p$ using the witness $(\text{sig}, r)$; in the latter case (secret key $sig\_k_\alpha$, associated with a public key generated by $\mathcal{A}$ during the preprocessing stage, has been already extracted), the protocol $\Pi_p$ is executed as in the previous case by using $sig\_k_\alpha$ to computes a witness;

4. upon receiving a message generated as in '*P step 1*' *(right interaction) on input $x, pk$:* $\mathcal{E}$ sends to $\mathcal{A}$ a message computed as in '*V step 2*'; here, the witness used by $\mathcal{E}$ to successfully complete the subprotocol $\Pi_v$ is the secret key $sig\_k_b$ that $\mathcal{E}$ generated in the preprocessing stage;

5. upon receiving a message generated as in '*V step 2*' *(left interaction) on input $y, pk'$:* $\mathcal{E}$ checks whether the obtained transcript of subprotocol $\Pi_v$ is accepting and $pk' \neq pk$; if so, it runs the extractor associated with the $\Sigma$-protocol $\Pi_v$ to extract the secret key $sig\_k_\alpha$ associated with $pk'$; once this extraction is completed, $\mathcal{E}$ rewinds $\mathcal{A}$ until right before the message generated in '*P step 1*' is sent to $\mathcal{A}$, and runs a modified version of '*P step 1*', where the modification is exactly as done in item 3 above;

6. upon receiving a message generated as in '*P step 2*' *(right interaction) on input $x, pk$:* $\mathcal{E}$ checks whether the transcript of subprotocol $\Pi_p$ in this right session is accepting; if so, it runs the extractor (using the rewinding technique) for the $\Sigma$-protocol $\Pi_p$, returns the entire conversation $extr_{right}(x)$ between $\mathcal{A}$ and the extractor for the $\Sigma$-protocol $\Pi_p$ and halts.

The predicate $\rho$ is simply defined as the algorithm that, on input $x_i, pk, extr_{right}(x_i)$ (with respect to the $i$th right session), writes $extr_{right}(x_i)$ as a pair of conversations of the 4-round $\Sigma$-protocol $\Pi_p$; if both conversations are accepting, $\rho$ can compute a witness for '$x_i \in L$', by using the properties of the extractor for $\Sigma$-protocols.

We start by noting some facts about the above construction of $\mathcal{E}, \rho$ and later prove that they satisfy Definition 2.2.

First of all, we note that $\mathcal{E}$ successfully simulates $\mathcal{A}$'s view in the right interactions; this can be seen by noting that in items 2 and 4 of $\mathcal{E}$'s construction, the next message for $\mathcal{A}$ is generated identically as in $V$'s algorithm.

Furthermore, we note that $\mathcal{E}$'s simulation of $\mathcal{A}$'s view in the left interactions only differs from $\mathcal{A}$'s view in a real execution of $\mathcal{A}$'s attack in the following: instead of the witness $w_i$ for $y_i \in L$, the secret key is used to compute a signature that is the witness used in subprotocol $\Pi_p$. We will show that this difference is computationally indistinguishable by $\mathcal{A}$, or otherwise we can use $\mathcal{A}$ to violate the witness-indistinguishability of $\Pi_v$ or the binding property of the commitment scheme used. To prove this, we construct a non-uniform hybrid simulator taking as input all these secret keys and all the witnesses of statements in the left interactions, which computes commitments exactly as $\mathcal{E}$ does and executes $\Pi_p$ using the same witness of the statement used by the honest prover. It is easy to see that the transcript generated by the hybrid simulator is indistinguishable from both the one in real interactions and the one generated in the extraction. Now we can claim that $\mathcal{E}$, running in expected polynomial time, can output a valid witness $w'$ with probability negligibly close to $p^i_\mathcal{A}(x_i)$. Then, if $\mathcal{E}$ extract the witness $w'$ successfully, one of the following two events must occur:

Event 1: $(x_i, w') \in R_L$;
Event 2: $w' = (\text{sig}', r), c = COM(\text{sig}', r))$ and $\text{sig}'$ is a valid signature of $vk'$ corresponding to $ver\_k_0$ or $ver\_k_1$.

If we prove that Event 2 occurs with negligible probability, the proof of Theorem 3.1 is complete. Then, in the rest of the proof we concentrate on proving that event 2 occurs with negligible probability, or otherwise we can contradict at least one of these three properties: the unforgeability of the signature scheme, the binding of the commitment scheme, or the witness-indistinguishability of the $\Sigma$-protocol used.

We start by assuming (towards contradiction) that the probability that Event 2 occurs is not negligible, and we construct a non-uniform algorithm $\mathcal{B}$ with access to a signing oracle that plays similarly to the extractor $\mathcal{E}$ and attempts to violate the security against adaptive chosen message attack of scheme $(KG_{1-b}, Sig_{1-b}, Ver_{1-b})$.

The non-uniform algorithm $\mathcal{B}$ is given access to the signing oracle of the signature scheme $(KG_{1-b}, Sig_{1-b}, Ver_{1-b})$, taking as auxiliary input $pk = (ver\_k_0, ver\_k_1, sig\_k_b)$ generated by the extractor $\mathcal{E}$, invokes the adversary $\mathcal{A}$, feeds it with the public key $(ver\_k_0, ver\_k_1)$ and works in exactly the same way as $\mathcal{E}$. Note that in this extraction process executed by $\mathcal{B}$, When $\mathcal{A}$ start a session under the public key generated by $\mathcal{E}$ (i.e., $pk$) in left interactions and $\mathcal{B}$ flips a bit $j \neq b$, $\mathcal{B}$ can obtain the signature on a verification key $vk'$ of the one-time signature scheme (generated by itself) by querying the signing oracle of $(KG_{1-b}, Sig_{1-b}, Ver_{1-b})$. We then denote as $w'_\mathcal{B}$ the witness extracted by $\mathcal{B}$ at the end of its execution, and as $vk'_\mathcal{B}$ the public key associated to the $i$-th right interaction, the latter being defined as the interaction where $\mathcal{B}$ is successful in rewinding the adversary $\mathcal{A}$, obtaining two different transcript of $\Pi_p$, and thus computing $w'_\mathcal{B}$.

Note that $\mathcal{E}$ and $\mathcal{B}$ implicitly define two extraction procedures, to which we will hereon refer as $Ext_\mathcal{E}$ and $Ext_\mathcal{B}$. We note that $\mathcal{A}$'s view is the same in both $Ext_\mathcal{E}$ and $Ext_\mathcal{B}$, and therefore event 2 occurs in $Ext_\mathcal{B}$ with exactly the same probability as that of Event 2 occurring in $Ext_\mathcal{E}$.

The proof continues with the following two claims.

*Claim 1.* Let $p$ be the probability that Event 2 occurs in $Ext_B$, and $q$ be the probability that the algorithm $\mathcal{B}$ outputs a signature $\text{sig}'_\mathcal{B}$ (contained in $w'_\mathcal{B}$) such that $(Ver_{1-b}(ver\_k_{1-b}, \text{sig}'_\mathcal{B}, vk'_\mathcal{B}) = 1$. If $p$ is non-negligible, then so is $q$.

Assume, towards contradiction, that Claim 1 is false; that is, $p$ is non-negligible, and $q$ is negligible. We now show that this contradicts either the witness indistinguishability of $\Pi_v$ or the computational binding of the commitment scheme $COM$.

We first note that during the above extraction procedures $Ext_\mathcal{B}$ and $Ext_\mathcal{E}$, the view of $\mathcal{A}$ in the left interactions is independent of $b$ (Note that in the executions of a left session under the public key $pk$, $\mathcal{E}$ or $\mathcal{B}$ randomly choose one of two secret keys associated with $pk$ and use this secret key to compute a witness). If the probability $q$ is negligible, this implies that if $\mathcal{B}$ always uses the witness $sig\_k_b$ to execute $\Pi_v$ during the extraction (i.e, $Ext_\mathcal{B}$), $\mathcal{B}$ always outputs a signature corresponding to $ver\_k_b$ except with negligible probability. Let $l$ be the number of right sessions[1] executed before the end of session $i$; by standard hybrid arguments, there must be $j \in \{1, ..., i\}$ such that if $\mathcal{B}$ uses the witness $sig\_k_b$ to execute $\Pi_v$ in $j$th right session, it will output a valid signature under $ver\_k_b$ with probability at least $(p-q)/l$ (note that the probability that it outputs a valid signature under $ver\_k_{1-b}$ is still negligible). It is easy to see that if the $j$-th proof (i.e., the $j$th execution of $\pi_v$) in the right interactions has been completed before '$P$ Step 1' in the $i$-th right session, $\mathcal{A}$ can be used to break the property of witness indistinguishability of $\Pi_v$ (it distinguished which witness used in the $j$th proof) because during the extraction we do not rewinds the $j$th right proof[2]. If the $j$-th proof in the right session completed after the $P$ Step 1 in the $i$-th session, we can construct a non-uniform algorithm $\mathcal{B}'$ to break the computational binding of the commitment scheme $COM$.

The non-uniform algorithm $\mathcal{B}'$ takes as auxiliary input $(ver\_k_0, ver\_k_1, sig\_k_0, sig\_k_1)$ (with input both signing keys), feeds $\mathcal{A}$ with the public key $(ver\_k_0, ver\_k_1)$ and runs $\mathcal{A}$'s key generation algorithm to get all public keys of $\mathcal{A}$. Then it performs the following extraction:

1. Simulation: acts exactly like $\mathcal{E}$ in left simulation in extraction. For the simulation in right interactions, $\mathcal{B}'$ picks a random bit $b$ and uses $sig\_k_b$ as witness in all right proofs (i.e., the right executions of $\Pi_v$) until it received the message $a$ in $P$ Step 1 in the $i$-th right session (Let $f$ be the first message of $\Pi_p$ in this session). After received $a$, $\mathcal{B}'$ continues independently in the two following games;

2. Game 0: $\mathcal{B}'$ uses $sig\_k_0$ as witness in all right proofs (including the $j$th proof) that completed after $P$ Step 1 in the $i$th right session to end the whole extraction. After $\mathcal{B}'$ obtains a accepting transcript $(f, a, e_0, z_0)$ of $\Pi_p$ in the $i$-th right session, it rewinds to the point of beginning of $V$ Step 2 in $i$-th session in the right interactions and replays this step by sending another random challenge $e'_0 \neq e_0$ until he gets another accepting transcript $f, a, e'_0, z'_0$ of $\Pi_p$.

---
[1]When $\mathcal{B}$ rewinds $\mathcal{A}$ to get his secret keys during the left extraction in $Ext_\mathcal{B}$, $\mathcal{A}$ may rewind $\mathcal{B}$ in right interactions and this may result in some new right sessions. The "right sessions" mentioned here do NOT include this kind of right sessions (generated during left extractions).

[2]Indeed, the $j$th (right) proof may be rewound by $\mathcal{A}$ due to left extraction (in order to get the adversary's secret key). However, we note that the transcript during the left extraction does not appear in $\mathcal{A}$'s view because after rewinding $\mathcal{A}$ each time $\mathcal{B}$ resets $\mathcal{A}$ to its state just before the rewinding (so the $\mathcal{A}$'s memory of the conversation during left extraction is "deleted"). In another words, if the $j$th proof (given by the verifier) in the right interactions has been completed before the '$P$ Step 1' in the $i$-th right session, $\mathcal{A}$ does not see any rewinding of the $j$th right proof.

3. Game 1: repeats Rewinding Game 0 twice and obtains two accepting transcript $(f, a, e_1, z_1)$ and $(f, a, e'_1, z'_1)$ of $\Pi_p$, but in this game $\mathcal{B}'$ uses $sig\_k_1$ as witness in all right proofs (including the $j$th proof) that completed after $P$ Step 1 in the $i$th right session to end the whole extraction. Since the $\Pi_v$ is *partial-witness-independent*, and note that a proof that completed after $P$ Step 1 in the $i$th right session means $\mathcal{A}$ has seen at most the first message of $V$ (i.e., V step 1) in that proof, so $\mathcal{B}'$ can choose different witness to complete those right proofs after $\mathcal{A}$ sent the first message $a$ of $\Pi_p$ ($P$ Step 1) in the $i$th right session.

Note that the message $a$ sent by $\mathcal{A}$ in the $i$th right session contains a commitment $C$ and a verification key $vk'$ of a one-time strong signature scheme. Clearly, with probability negligibly close to $p^2$, $\mathcal{B}$ will output two valid witness $w'_0 = (\texttt{sig}'_0, r_0)$ and $w'_1 = (\texttt{sig}'_1, r_1)$ from the above two games such that the following holds: $\texttt{sig}'_0 = Sig_0(sig\_k_0, vk')$, $\texttt{sig}'_1 = Sig_1(sig\_k_1, vk')$, $c = COM(\texttt{sig}'_0, r_0))$ and $c = COM(\texttt{sig}'_1, r_1))$. This contradicts the computational-binding property of the scheme $COM$.

In sum, if the event 2 occurs in $Ext_E$ with a non-negligibly probability $p$, the algorithm $\mathcal{B}$ can output a signature $\texttt{sig}'_{\mathcal{B}}$ of $vk'_{\mathcal{B}}$ such that $(Ver_{1-b}(ver\_k_{1-b}, \texttt{sig}'_{\mathcal{B}}, vk'_{\mathcal{B}}) = 1$ with a non-negligible probability. If $\mathcal{B}$ did not query the signing oracle on the message $vk'_{\mathcal{B}}$, it breaks the scheme $(KG_{1-b}, Sig_{1-b}, Ver_{1-b})$.

*Claim 2.* $vk'_{\mathcal{B}}$ is different from any verification key of the one-time strong signature scheme $(KG', Sig', Ver')$ generated by $\mathcal{B}$ in the left interactions.

Assume otherwise, $vk'_{\mathcal{B}}$ appeared in the transcript of $j$th left session for some $1 \leq j \leq s$, which we denote with $tr^j_{left}$ and $tr^j_{left} = (tran_j, \delta_j)$ according to our notation. Let $tr^i_{right} = (tran_i, \delta_i)$ be the accepting transcript in the $i$th left session before $\mathcal{B}$ rewinds. Then, we have $Ver'(vk'_{\mathcal{B}}, tran_j, \delta_j) = 1$, $Ver'(vk'_{\mathcal{B}}, tran_i, \delta_i) = 1$ and $tr^j_{left} \neq tr^i_{left}$ (according to the definition 3), thus one the following two cases must occur: 1) $tran^j \neq tran^i$, this means $\mathcal{A}$ produced a valid signature on a different message; 2) $\delta_j \neq \delta_i$, this means $\mathcal{A}$ produced a valid different signature (on a possibly different message). It is clear that in both cases $\mathcal{A}$ violates the security requirements of a one-time strong signature scheme. $\square$

### 3.1 Efficient Instantiations of Our Transformation

We present two efficient instantiations based on Pedersen's commitment scheme (based on DL assumption)[35] and Elgamal commitment scheme (based on DDH assumption) respectively. For the signature schemes, we employ the Boneh and Boyen's short signature scheme (based on Strong DH Assumption)[6] and the one-time strong signature scheme in [19]. We first note that when we make the assumptions underlying these specific schemes, our protocols $\Pi_p$ and $\Pi_v$ can be reduced into 3-round protocols, and $\Pi_v$ still enjoys the partially-witness-independent property. We now briefly recall these schemes.

*The Boneh and Boyen's scheme.* This scheme is based on bilinear map and its security relies on the strong Diffie-Hellman Assumption. Let $G_1$ and $G_2$ are two cyclic groups of prime order $q$ with generators $g_1, g_2$ respectively, $\psi$ is an efficiently computable isomorphism $\psi : G_2 \to G_1$ with $\psi(g_2) = g_1$. We say $(G_1, G_2)$ are bilinear groups if there exist a group $G_T$, an isomorphism $\psi$ defined above, and a bilinear map $\epsilon : G_1 \times G_2 \to G_T$ such that for all $u \in G_1, v \in G_2$ and $a, b \in \mathbb{Z}$, $\epsilon(u^a, v^b) = \epsilon(u, v)^{ab}$ and $\epsilon(g_1, g_2) \neq 1$.

The signature scheme operates as follows: 1) **Key generation algorithm** $KG$. Pick a random generator $g_2 \leftarrow_R G_2$ and set $g_1 = \psi(g_2)$. Pick random $x, y \leftarrow_R \mathbb{Z}_q^*$, and compute $u \leftarrow g_2^x \in G_2$ and $v \leftarrow g_2^y \in G_2$. Also compute $z \leftarrow \epsilon(g_1, g_2) \in G_T$. The verification key is $(g_1, g_2, u, v, z)$, the signing key is $(x, y)$. 2) **Signing algorithm** $Sig$. Given a signing key $x, y \in \mathbb{Z}_q^*$ and a message $m \in \mathbb{Z}_q^*$, pick a random $r \leftarrow_R \mathbb{Z}_q^*$ and compute $\sigma \leftarrow g_1^{1/(x+m+yr)} \in G_1$. Here $1/(x+m+yr)$ is computed modulo $q$. In the unlikely event that $x+m+yr = 0$ we try again with a different random $r$. The signature is $(\sigma, r)$. 3) **Verification algorithm** $Ver$. Given a verification key $(g_1, g_2, u, v, z)$, a message $m \in \mathbb{Z}_q^*$, and a signature $(\sigma, r)$, verify $\epsilon(\sigma, u \cdot g_2^m \cdot v^r) = z$, if this is the case, output *valid*, otherwise output *invalid*.

*Pedersen's commitment scheme.* Let $p$, $q$ be two primes such that $p = 2q + 1, |q| = n$, where $n$ is the security parameter, and let $G_q$ denote the subgroup of $\mathbb{Z}_p^*$ with order $q$ in which the DL problem is hard and $g$ is a generator of the subgroup. To commit a value, the receiver chooses a random numbers $x \in \mathbb{Z}_q$ first (this results in the verifier sending this random value in $V$ step 1 in the protocol depicted in figure 1), computing $h = g^x$, and sends $h$ to the sender. Then the sender $S$ commits to a value $y$ as follows: it randomly chooses $r \in \mathbb{Z}_q$, computing $C = g^y h^r$, send $C$ to the receiver. To decommit a commitment $C$, the sender delivers $y$ and $r$. The binding property of this commitment scheme lies in DL assumption on the subgroup $G_q$. We note that this scheme enjoys perfect-hiding property: the distribution of the commitments is indistinguishable for all powerful receiver $R^*$.

*Elgamal commitment scheme.* It is a basic application of Elgamal encryption scheme. Let $p, q, g, G_q$ as described above, but in this scheme we assumes that the DDH problem in $G_q$ is hard. To commit a value $y$, the sender $S$ chooses a random numbers $x \in \mathbb{Z}_q$ and computes $h = g^x$ (note that the sender chooses $x$ itself and the committing stage does not require interaction), then it commits to a value $y$ as follows: it randomly chooses $r \in \mathbb{Z}_q$, computing $C = (g^r, g^y h^r)$, send $C$ to the receiver. To decommit a commitment $c$, the sender delivers $y$ and $r$. The hiding property of this commitment scheme lies in DDH assumption on the subgroup $G_q$, and this scheme enjoys perfect-binding property: it is impossible to open a commitment $C$ in different way for all powerful receiver $S^*$.

In the following specific instantiations, we use the same prime $q$ in both the Boneh and Boyen's scheme and the commitment schemes, therefore the length of the each part of the signature $(\sigma, r)$ is $\lceil \log_2 q \rceil$. Let $\lceil \sigma \rceil$ be the integer representation of $\sigma$, clearly $\lceil \sigma \rceil$ is in $\mathbb{Z}_q$ which is the message space of the Pedersen's scheme. We assume without loss of generality that the length of the public key $vk'$ of the efficient one-time strong signature scheme is $\lceil \log_2 q \rceil$, the message size required by the the Boneh and Boyen's scheme (in fact, we can sign arbitrary message by hashing it first without loss of security, see [6]).

It is easy to instantiate the protocol $\Pi_v$ in which the verifier prove the knowledge of the secret key of the The Boneh and Boyen's signature scheme by using the well-known proof of knowledge of one out of two discrete logarithms. Now, consider the following two statements, where the common input consists of a commitment key $C$, signature verification keys generated according to the key-generation algorithm of the Boneh and Boyen's schemes and a message $m$ (i.e, the verification key of the one time signature scheme):

1. There exist $\sigma$, $r$ such that $C = COM(\sigma, r)$ and $\sigma$ is a valid signature of $m$ under one of two signature verification keys, according to Boneh and Boyen's signature verification algorithm.
2. There exist $\sigma, r$ such that $C = COM(\sigma, r)$ and $\sigma$ is a valid signature of $m$ under a given signature verification key, according to Boneh and Boyen's signature scheme.

We observe that if we can design an efficient $\Sigma$-protocol for statement 2, then we can apply the OR-composition technique discussed in Appendix **??**, and obtain an efficient $\Sigma$-protocol for statement 1. Therefore, we focus on giving an efficient instantiation of a $\Sigma$-protocol for statement 1. (We stress that the $\Sigma$-protocol for statement 2 based on El Gamal's commitment scheme satisfies only special honest-verifier *computational* zero-knowledge, and so does the corresponding $\Sigma$-protocol for the statement 1.)

The $\Sigma$-protocols for the statement 1 based on Pedersen's commitment scheme and the Elgamal commitment scheme are depicted in Figure 2 and Figure 3, respectively. The common input consists of two commitments $C_1$, $C_2$ to the signature $(\lceil \sigma \rceil, r)$ (where $C_1$ is a commitment to $\lceil \sigma \rceil$ and $C_2$ is a commitment to $r$), the verification key $(g_1, g_2, u, v, z)$ of the Boneh and Boyen's scheme, parameters $g, h, p, q$ of Pedersen's scheme and message $m$.

*Ideas behind the $\Sigma$-protocols for statement 1.* Our proof system combines the following two subprotocols: (1) a proof of knowledge of the value committed to by the commitment key (for Pedersen's scheme, the proof system of this statement is proposed in [32]); (2) a proof that the value contained in the commitment is a vlid signature on a known massage. This can be done as follows: first the prover generates $A$ and $F$ (see figure 2 and figure 3) on which the verifier will check whether these element pass a randomized variant of the signature verification equation (we note that it is possible to see that sending both $F$ and $t$ will not harm the prover, as Boneh and Boyen's signatures are properly randomized). Then the prover sends three specific element $B$, $D$ and $E$ whose purpose is to show that the signature determined by $F$ and $V$ matches the value committed to in the commitment.

For clarity of description, we omit specifying the groups over which the exponentiations are performed, as they are clear from the context.

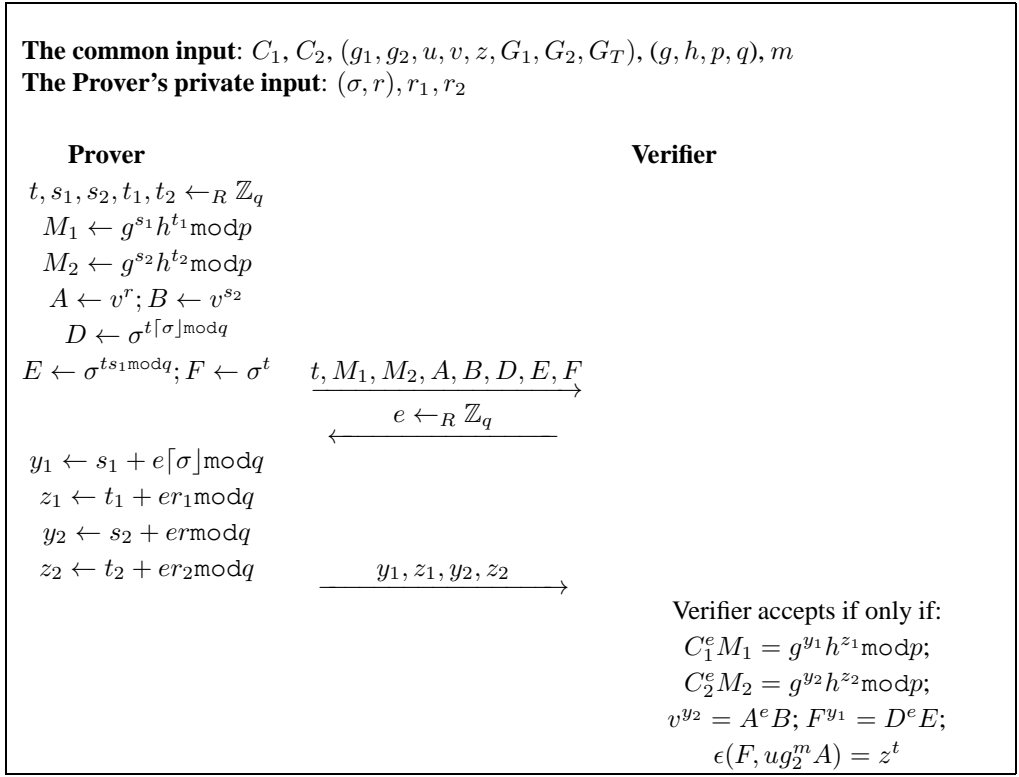

**Figure 2.** The $\Sigma$-protocol for the statement 1 based on Pedersen's commitment scheme, i.e., proof of relationship
$\{((C_1, C_2, g_1, g_2, u, v, z, g, h, p, q, m); (\sigma, r, r_1, r_2)) \mid C_1 = g^{\lceil \sigma \rceil} h^{r_1} \wedge C_2 = g^r h^{r_2} \wedge \epsilon(\sigma, u g_2^m v^r) = z\}$.

For the $\Sigma$-protocol based on Pedersen's commitment scheme, it is easy to see that we can compute the witness from two different transcripts $(a, e, z)$ and $a, e', z'$ of two executions of above protocol. Special honest-verifier zero-knowledge is also clear: given a challenge $e$, a simulator randomly chooses $s \leftarrow_R \mathbb{Z}_q^*$ and computes $\sigma \leftarrow g_1^s \in G_1$, picks $t \leftarrow_R \mathbb{Z}_q$, $y_1 \leftarrow_R \mathbb{Z}_q$, $y_2 \leftarrow_R \mathbb{Z}_q$, $z_1 \leftarrow_R \mathbb{Z}_q$ and $z_2 \leftarrow_R \mathbb{Z}_q$; then, it computes $M_1 \leftarrow g^{y_1} h^{z_1} C_1^{-e} \bmod p$, $M_2 \leftarrow g^{y_2} h^{z_2} C_2^{-e} \bmod p$, $F \leftarrow \sigma^t \in G_1$, $D \leftarrow \sigma^{t \lceil \sigma \rceil \bmod q} \in G_1$, $E \leftarrow D^e F^{-y_1} \in G_1$, $A \leftarrow u^{-1} g_2^{-m} g_2^{1/s} \in G_2$, $B \leftarrow v^{y_2} A^{-e} \in G_2$, and it is easy to check that the "transcript" generated in this way is with the same distribution of the real execution. Note that the Boneh and Boyen's scheme requires $(x + m + yr) \in \mathbb{Z}_q^*$, and this is why the simulator can randomly choose $s$ from $\mathbb{Z}_q^*$.

We can analyze the $\Sigma$-protocol based on Elgamal commitment scheme in similar way, but this protocol enjoys only special honest-verifier *computational* zero-knowledge.

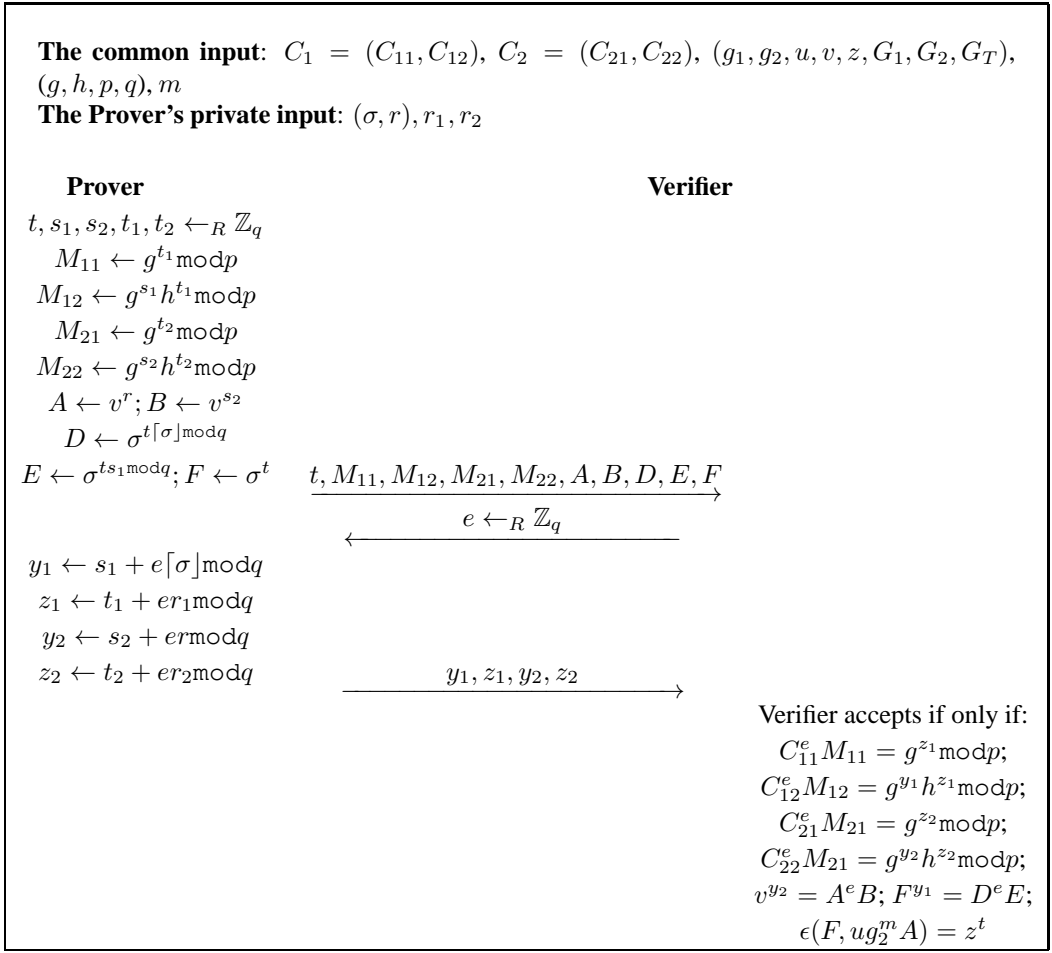

**Figure 3.** The $\Sigma$-protocol for statement 1 based on Elgamal commitment scheme, i.e, proof of relationship
$\{((C_{11}, C_{12}, C_{21}, C_{22}, g_1, g_2, u, v, z, g, h, p, q, m); (\sigma, r, r_1, r_2)) \mid (C_{11}, C_{12}) = (g^{r_1}, g^{\lceil\sigma\rfloor}h^{r_1}) \wedge (C_{21}, C_{22}) = (g^{r_2}, g^r h^{r_2}) \wedge \epsilon(\sigma, ug_2^m v^r) = z\}$.

## 4 Impossibility Result for Concurrently Non-Malleable Zero-Knowledge in the BPK model

In this section we ask the natural question of whether it is possible to further reduce the setup assumptions used in Section 3 for the construction of concurrently non-malleable zero-knowledge arguments of knowledge. We consider the setup assumptions of the BPK model. Due to the lack of authenticated access to the public file (as in the APK model), no assumption is made in the BPK model about guaranteeing that the public file is correctly read from provers at any time during the proof stage. In other words, if we consider (without loss of generality) any prover's action of reading the public file as a message exchange between this prover and the public file site, we allow the communication between the prover and the public file site to be unauthenticated, and thus subject to modification by a man-in-the-middle adversary. As a consequence, we obtain the following negative result:

**Theorem 4.1** Let $L$ be a language and let $R_L$ be the relation associated to $L$. If in the BPK model, there exists a concurrently non-malleable zero-knowledge arguments of knowledge for $R_L$, then $L$ is in BPP.

As the APK model can be seen as a minor strengthening of the BPK model, this negative result seems to imply that the setup assumptions that are sufficient in Theorem 3.1 to obtain concurrently non-malleable zero-knowledge arguments of knowledge for relations associated with non-trivial languages, are essentially optimal. In other words, the above result says that authenticated communication between all provers and the public file site is a necessary assumption to obtain these protocols. The proof of Theorem 4.1 is obtained by transforming a concurrently non-malleable zero-knowledge arguments of knowledge for $R_L$ in the BPK model into one for the same relation in the standard model (that is, without setup assumption), and then invoking the result of [29], which says that such protocols, even regardless of their (polynomial) round complexity, are only possible for relations associated with trivial languages in the standard model.

*proof of theorem 4.1.* The proof is obtained by transforming a concurrently non-malleable zero-knowledge arguments of knowledge for $R_L$ in the BPK model into one for the same relation in the standard model (that is, without setup assumption), and then invoking the result of [29], which says that such protocols, even regardless of their (polynomial) round complexity or simulation paradigm (black-box or non-black-box), are only possible for relations associated with trivial languages in the standard model.

Let, $(P, V)$ be a concurrently non-malleable zero-knowledge arguments of knowledge for $R_L$ in the BPK model; here, $V = (V_1, V_2)$, where $V_1$ runs in the preprocessing stage and $V_2$ runs in the proof stage.

We define a protocol $(P', V')$ in the standard model as follows. $V'$ divides its random tape in two parts (of appropriate length) as $r_1 || r_2$; then, it first runs $V_1$ on input the security parameter $1^n$ and using $r_1$ as randomness, thus obtaining $(pk, sk)$ and sending $pk$ to $P'$; then it continues running $V_2$ using $r_2$ as randomness. $P'$ is almost identical to $P$, the only difference being that whenever $P$ uses the value $pk_V$ from the public file, $P'$ uses the value $pk$ received from $V'$. Here, we assume without loss of generality that $P$ sends the first message in the execution of $(P, V)$, if not, we should add another $P'$ step: after receiving the public key $pk$ from $V'$, $P'$ replies with a message 'I received $pk$' and then both parties continue as discussed above (i.e., $V'$ runs $V_2$ and $P'$ runs $P$).

We now need to show that $(P', V')$ is a concurrently non-malleable zero-knowledge argument of knowledge in the standard model, under the assumption that $(P, V)$ is a concurrently non-malleable zero-knowledge argument of knowledge in the BPK model. To that purpose, we first show how to formally define the notion of concurrently non-malleable zero-knowledge argument of knowledge in the BPK model, and recall the definition of concurrently non-malleable zero-knowledge argument of knowledge in the standard model.

*The definition in the BPK model.* We note that concurrently non-malleable zero-knowledge argument of knowledge in the BPK model can be formally defined almost identically as in the APK model, the only differences being the following. First, the equality $F_p = F$ does not hold any more in the BPK model; that is, each prover $P$'s access at time $t$ to the public file $F$ may return a file $F_{P,t}$ that may not be equal to $F$ due to adversarial action from a man-in-the-middle (Though each prover can keep the file in local memory, The adversary still can invoke different provers in the left interactions, and these provers do not aware the existence of each other, thus the public file kept by a prover may be different form the one kept by another prover). Second, in modeling the adversary $\mathcal{A}$'s man-in-the-middle attack, for sake of generality we consider the worst possible behavior from $\mathcal{A}$, where each prover $P$'s access at time $t$ to the public file $F$ may return a file $F_{P,t,\mathcal{A}}$ that is arbitrarily chosen by $\mathcal{A}$. The requirements of completeness, concurrent zero-knowledge, extraction and simulatability of $\mathcal{A}$'s view are consequently modified.

*The definition in the standard model.* We note that concurrently non-malleable zero-knowledge argument of knowledge in the standard model were first defined in [17], where the extraction property says that for each adversary who plays as a man-in-the-middle and is given access to a polynomial number of provers in left interactions, there exists a simulator that is not given access to such provers and is essentially as successful in making the verifier in the right interaction accepts. In the same paper, it was also showed that the property guaranteeing extraction of a valid witness from the adversary implies the concurrent non-malleability property of the given protocol.

*Rest of the proof (sketch).* The completeness property of $(P', V')$ immediately follows from the analogue property of $(P, V)$.

To see that $(P', V')$ satisfies the concurrent zero-knowledge property, we construct a simulator $S$ for any probabilistic polynomial time $V'^*$. Let $S = \mathcal{M} = [\mathcal{M}_P, \mathcal{M}_V]$ (the latter algorithms are guaranteed by the assumed simulatability of $\mathcal{A}$'s view property of protocol $(P, V)$). Then $S$ acts as following: $\mathcal{M}_V$ does nothing; whenever $V'^*$ sends a arbitrary public key $pk$ to $\mathcal{M}_P$, $\mathcal{M}_P$ starts a new session under $pk$ as it interacts with the man-in-the-middle adversary $\mathcal{A}$ (here, $\mathcal{M}_P$ treats $V'^*$ as $\mathcal{A}$); at the end of simulation, $S$ outputs the transcript of all interactions. This simulator works because in the (unauthenticated) BPK model, $\mathcal{A}$ actually delivers an arbitrary public key to the prover before each session (again, note that $\mathcal{A}$ can invoke different provers in every left session, so prover's memory of the public file does not keep $\mathcal{A}$ from delivering an arbitrary public key to a prover before each session) in the left interactions, and thus we can view the interactions between $V'^*$ with several honest provers (in the protocol $(P', V')$) as the left interactions in which $V'^*$ plays the man-in-the-middle adversary interacting with several honest provers and verifiers with respect to protocol $(P, V)$ (indeed $V'^*$ does not interacts with $V$).

The protocol $(P', V')$ also enjoys the extraction property (and therefore the concurrent non-malleability property). We can construct an extractor $E'$ for the man-in-the-middle adversary $\mathcal{A}'$ with respect to $(P', V')$ by modifying the extractor $E$ (this guaranteed by the extraction property of $(P, V)$) for the adversary $\mathcal{A}$ with respect to $(P, V)$ in this way: instead of generating a public file (consisting of a number of public keys) and feeding this file to $\mathcal{A}$ first (this can be seen to be an implicit requirement for $E$ or otherwise the extraction property would not hold for $(P, V)$), $E'$ stores this file itself. Whenever $\mathcal{A}'$ want to interact with $V'_i$, $E'$ first sends the public key in the $i$-th entry of the public file, and then acts just identically to $E$. We can claim that if $E'$ does not work, so $E$ doesn't, because by modifying $\mathcal{A}'$, we can easily construct an adversary $\mathcal{A}$ which will break the extraction property of $(P, V)$.